\documentclass[a4paper,11pt]{article}
\usepackage{graphicx,amssymb,amstext,amsmath,mathabx,mathtools,esvect,xparse, bbm}
\usepackage{xcolor}
\usepackage{floatflt}
\usepackage{microtype}

\definecolor{cbl}{rgb}{0,0,1}               % bleu\

\newcommand{\bc}{\begin{center}}
\newcommand{\ec}{\end{center}}
\def\ba#1{\begin{array}{#1}\displaystyle}
\newcommand{\ea}{\end{array}}

\newcommand{\beq}{\begin{equation}}
\newcommand{\eeq}{\end{equation}}
\newcommand{\beqa}{\begin{eqnarray}}
\newcommand{\eeqa}{\end{eqnarray}}

\newcommand{\bi}{\begin{itemize}}
\newcommand{\ei}{\end{itemize}}
\newcommand{\bra}{\langle}
\newcommand{\ket}{\rangle}
\newcommand{\Z}{{\mathbb{Z}}}

\newcommand{\dd}{\mathrm{d}}
\newcommand{\ii}{{\mathrm{i}}}

\newcommand{\TT}{{\cal T}}

\newcommand{\he}{\color{red} \heartsuit}
\newcommand{\club}{\color{blue} \clubsuit}

\newcommand{\spa}{\color{teal} \spadesuit}

\usepackage{cite}

\definecolor{purple_nice}{rgb}{0.4,0.2,0.7}
\definecolor{fuel_blue}{RGB}{42,162,185}
\definecolor{YInMn_blue}{RGB}{46, 80, 144}
\definecolor{ultramarine}{RGB}{63, 0, 255}
\definecolor{KLEIN_blue}{rgb}{0, 0.18, 0.65}
\usepackage[linktocpage=true]{hyperref}
\hypersetup{
    colorlinks=true,
    linkcolor=YInMn_blue,
    citecolor=ultramarine,
    filecolor=fuel_blue,
    urlcolor=KLEIN_blue,
}
\makeatletter
\renewenvironment{abstract}{%
      \begin{center}%
        {\bfseries \normalsize\abstractname\vspace{\z@}}%  %% <- here I've added \Large
      \end{center}%
      \quotation}
    {\endquotation}
\makeatother
\usepackage[normalem]{ulem}  % FOR THE \sout COMMAND

\newcommand{\fed}[1]{{\color{black}  #1}}

\begin{document}

\begin{titlepage}
\title{Time Evolution of the Symmetry Resolved\\ Entanglement Entropy after a Mass Quench}
\author{Federico Rottoli${}^{\spa}$\footnote{Corresponding author}, Michele Mazzoni${}^\diamondsuit$,  Fabio Sailis${}^{\club}$ and Olalla A. Castro-Alvaredo${}^{\he}$\\[0.3cm]}
\date{\small 
{${}^{\spa}$}  Dipartimento di Fisica dell'Universit\`a di Pisa and INFN Sezione di Pisa,\\ Largo Pontecorvo 3, 56127 Pisa, Italy\\
\medskip
${}^{\diamondsuit}$ Dipartimento di Fisica e Astronomia, Universit\`a di Bologna and INFN\\
Sezione di Bologna, via Irnerio 46, 40126 Bologna, Italy\\
\medskip
${}^{\club \he}$ Department of Mathematics, City St George's, University of London,\\ 10 Northampton Square EC1V 0HB, UK
}
\maketitle
\begin{abstract}
In this paper we investigate the properties of the symmetry resolved entanglement entropy after a mass quench in the Ising field theory. Since the theory is free and the post-quench state known explicitly, the one-point function of the relevant (composite) branch point twist field can be computed using form factor techniques, similar to previous work on the branch point twist field and the magnetisation, respectively. We find that the symmetry resolved entropy grows linearly in time at the same rate as the total entropy, and that there are sub-leading oscillatory corrections. This result provides the first explicit computation of the out-of-equilibrium dynamics of the symmetry resolved entropy employing twist fields in quantum field theory and is consistent with existing results based on the quasiparticle picture.
\end{abstract}
\bigskip
\bigskip
\noindent {\bfseries Keywords:} Ising Field Theory,  Quantum Quench, Twist Fields, Entanglement
\vfill

\noindent 
${}^{\spa}$ federico.rottoli@df.unipi.it\\
${}^{\diamondsuit}$ 
    michele.mazzoni7@unibo.it\\
${}^{\club}$ fabio.sailis@city.ac.uk\\
${}^{\he}$ o.castro-alvaredo@city.ac.uk\\

\hfill \today
\end{titlepage}
\section{Introduction}
Over the  past decade there has been intense scientific activity in relation to both entanglement measures and out-of-equilibrium dynamics in many-body quantum systems, particularly in 1+1 dimensions and in the context of integrability and criticality, see e.g. \cite{specialissue, specialissue2, HolzheyLW94, Latorre1,Jin,Calabrese:2004eu, entropy, EEquench, ourhydro,theirhydro, Essler_2023, Doyon:2019nhl,BBDV} for some relevant reviews, special issues and key papers. The number of works investigating either of these problems and the intersection thereof is enormous. This letter provides one additional contribution to this ongoing discussion. We study the properties of the symmetry-resolved entanglement entropy (SREE) of the Ising field theory, following a mass quench. 

The SREE is a measure of entanglement that was originally proposed in \cite{GS} in the context of quantum field theory (QFT) and studied almost simultaneously in \cite{german3} for a quantum spin chain. A short review of its main applications and properties has been recently published \cite{ourreview}. A distinct feature is that this is a measure that exploits the presence of an internal symmetry in the theory under study in order to apportion the entanglement contribution to a specific symmetry sector. In other words, the total entanglement entropy can be expressed in terms of the contributions of each symmetry sector, with appropriate non-trivial weightings, and these individual contributions are the SREEs. In the Ising field theory, there is an internal discrete symmetry corresponding to the symmetry group $\mathbb{Z}_2$ so that there are two symmetry sectors and it is possible to define the SREEs associated to these sectors. The SREE and other related measures of free fermionic models at equilibrium have been studied in many papers such as \cite{Cornfeld:2018wbg,Bonsignori_2019,Capizzi_2020,Murciano:2021djk} for massless fermions and for free fermionic spin chains \cite{Bonsignori:2020laa,Ares:2022gjb,Ares:2022hdh}. There are also results for massive free fermions such as the Dirac free fermion, which is a free massive QFT with $U(1)$ symmetry \cite{Horvath:2021fks} and for the Ising field theory (Majorana fermion) with $\mathbb{Z}_2$ symmetry \cite{Horvath:2020vzs, CTFMichele}. Free fermions and their associated SREEs have also been studied in out-of-equilibrium situations \cite{Parez:2020vsp,Parez:2021pgq,Parez:2022xur}. An important result of these studies has been the observation that, following a quantum quench, the SREE grows linearly in time, similar to the total entropy, and that this is a natural consequence of the quasiparticle picture \cite{EEquench,AC}. 

Before going any further, it is useful to introduce some basic definitions so that we can more precisely explain what it is that we will be computing in the core part of the paper.  Let $|\Psi\ket$ be a pure state of a 1+1D QFT and let us define a bipartition of space into two complementary regions $A$ and $\bar{A}$ so that the Hilbert space of the theory $\mathcal{H}$ also decomposes into a direct product $\mathcal{H}_A \otimes \mathcal{H}_{\bar{A}}$. Then the reduced density matrix associated to subsystem $A$ is obtained by tracing out the degrees of freedom of subsystem $\bar{A}$ as
\beq
\rho_A=\mathrm{Tr}_{\bar{A}}(|\Psi\ket \bra \Psi|)\,.
\eeq
Most measures of entanglement are functions of this reduced density matrix. In particular, in the presence of an internal symmetry, we can also define a symmetry operator $Q$ and its projection onto subsystem $A$, $Q_A$. Assuming that $[Q_A, \rho_A]=0$ and taking $q$ to be the eigenvalue of operator $Q_A$ in a particular symmetry sector, then we can define the symmetry-resolved partition function ${\mathcal{Z}}_n(q)=\mathrm{Tr}_A(\rho_A^n \mathbb{P}(q))$ with $\mathbb{P}(q)$ the projector onto the symmetry sector of charge $q$. Then the SREEs are given by
\beq
S_n(q)=\frac{1}{1-n} \log\frac{{\mathcal{Z}}_n(q)}{{\mathcal{Z}}_1^n(q)}\quad \mathrm{and} \quad S(q)=\lim_{n\rightarrow 1} S_n(q)\,.
\label{sym}
\eeq
where $S_n(q)$ is the R\'enyi SREE and $S(q)$ the von Neumann SREE. The total R\'enyi and von Neumann entropies, denoted by $S_n$ and $S$ admit similar formulae with the projector $ \mathbb{P}(q)$ replaced by the identity operator. Similarly, $\mathcal{Z}_n=\mathrm{Tr}_A \rho_A^n$ can be interpreted as the (standard) partition function, {\fed{with $n$ the replica number or R\'enyi index}}. 

Indeed, a final important ingredient is the connection between these formal definitions and the kind of objects that are naturally computed in QFT, that is correlation functions of local fields. This connection can be established via the replica trick which identifies $\mathcal{Z}_n$ as the partition function of a replica theory defined in a multi-sheeted, cyclically-connected Riemann manifold (see \cite{entropy, ourreview} for good illustrations of this idea). The replica theory consists of $n$ identical copies of the model of interest, the Ising field theory, in our case. The connection to correlation functions is finally achieved when this partition function is identified with a correlator of symmetry fields associated with cyclic permutation symmetry of the copies. The field that implements this symmetry is known as branch point twist field \cite{entropy}. \fed{The partition function $\mathcal{Z}_n$ can be written as a correlation function of branch point twist fields, where the number of fields is the number of boundary points between entanglement regions}. If the QFT is integrable, then the form factor program \cite{smirnovbook,KW,entropy} provides a systematic approach to computing correlation functions, hence also the entropy. This methodology has been very successfully employed over the past 15 years (see e.g. \cite{entropy,next,nexttonext,Bianchini:2016mra,Castro-Alvaredo:2010scy,Castro-Alvaredo:2018dja}). More importantly for our purposes, it has been also generalised to composite branch point twist fields \cite{Horvath:2020vzs,Horvath:2021fks,CTFMichele}, that is, the type of branch point twist field \cite{CDL,Levi, BCDLR,bcd15} that is required in the context of symmetry resolution, as noted in \cite{GS}.

Let us then consider the branch point twist field $\TT$ and the composite branch point twist field $\TT_\mu$. These fields are characterised via their equal-time exchange relations with the fermion field of the Ising field theory. In the replica theory there are now $n$ such fermion fields denoted by $\Psi_j$ with $j=1,\ldots, n$ and $j+n\equiv j$ due to cyclic permutation symmetry. The equal-time exchange relations are in fact identical for $\TT$ and $\TT_\mu$ (and their hermitian conjugates) namely
\beq
\Psi_j (x)\TT(y)=\left\{ \begin{array}{cc}
\TT(y) \Psi_{j+1}(x)& y>x\\
\TT (y) \Psi_j(x) & y<x
\end{array}\right. \quad \mathrm{and}\quad \Psi_j (x)\TT^\dagger(y)=\left\{ \begin{array}{cc}
\TT^\dagger(y) \Psi_{j-1}(x)& y>x\\
\TT^\dagger(y) \Psi_j(x) & y<x
\end{array}\right.
\label{T}
\eeq 
and similarly for $\TT_\mu$ {\color{black} but with an additional minus sign in the second line of the equations above (see \cite{CTFMichele})}. The field $\TT_\mu$ can be interpreted as the composition of the fields $\TT$ and $\mu$, where $\mu$ is the so-called order field which implements $\mathbb{Z}_2$ symmetry in the Ising field theory\footnote{In the Ising field theory there are two symmetry fields associated with $\mathbb{Z}_2$ symmetry, usually called order and disorder field. In this paper the order field is $\mu$ and the disorder field is $\sigma$. In the disordered phase of the theory, the vacuum expectation value of $\mu$ is non-vanishing whereas it is vanishing for $\sigma$. The reverse is true in the ordered phase. This is the reason why in many papers the expectation value of $\sigma$ (ie. the magnetisation) is computed instead.}. In order to compute the SREE we will require the correlation functions of both $\TT$ and $\TT_\mu$ in a particular state. In this paper the state is determined by the out-of-equilibrium protocol that we are considering, that is, a mass quench. As we discus in the next section, considering the theory with fermion mass $m_0$ in the ground state and applying a subsequent time evolution characterised by the Hamiltonian of the same theory at mass $m$  gives rise to a post-quench state that is highly non-trivial. In such a state even the one point functions of $\TT$ and $\TT_\mu$ are rather intricate and we will focus on those. They are related to the SREE of a semi-infinite subsystem. 

Our study follows very closely the work \cite{IsingQuench} where the one-point function of the branch point twist field $\TT$ was studied in a lot of detail for the same quench. That study, in turn, was inspired by the work of Schuricht and Essler on the one-point function of the magnetisation \cite{SE}.
The same quench has also been considered in the lattice version of the present model, that is the XY model in a transverse magnetic field, where the leading scaling of the entanglement entropy can be computed analytically \cite{Fagotti_2008,CEF,CEF2}. For more general quenches, the entanglement entropy of free fermions following a quantum quench has been computed using ballistic fluctuation theory \cite{DDR}.

The main result of \cite{SE} and \cite{IsingQuench} was that both for the magnetisation and for the branch point twist field, the one-point function decays exponentially in time after a mass quench. {\fed{Exponential decay is observed in various settings, including in the context of cellular automata as shown for instance in \cite{Klobas:2024mlb}}}. However, because the theory is free and analytical methods are at our disposal, it is also possible in this case to describe many of the subleading corrections to this behaviour.  {\color{black} The formula obtained in \cite{IsingQuench} is 
\beqa
    \frac{_n\bra{\Omega}| \mathcal{T}(0, t) |{\Omega}\ket_n}{_n\bra{\Omega}|{\Omega}\ket_n} \approx\tilde{\tau}_n'  \exp{\left(- \frac{n\Gamma}{2} mt - \frac{n\alpha^2}{64 \pi mt}-\frac{\alpha \cos\frac{\pi}{2n}}{8 \sqrt{\pi } n \sin^2\frac{\pi }{2 n}}  \frac{\cos \left(2 m t-\frac{\pi}{4}\right)}{(mt)^{3/2}}\right)}\,,
    \label{formula2}
\eeqa 
where $\tilde{\tau}'_n$  is the post-quench ground state expectation value, $\Gamma, \alpha$ are functions of the pre- and post-quench masses which we will define below and $|\Omega\ket_n$ is the (non-normalised) pre-quench ground state in the $n$-replica theory. As we can see, exponential decay is the leading behaviour of the one-point function for large times. The entanglement entropy is proportional to the logarithm of the one-point function which leads to linear growth with time.  As we can see from the formula above, corrections are obtained which describe the dynamics also for relatively short times and scale as powers $t^{-a}$ with $a>0$ combined with oscillatory corrections of frequency $2m$, $m$ being the post-quench mass scale\footnote{These oscillatory corrections are a first order effect, from the viewpoint of perturbation theory on the ratio of the masses. For this reason this term can also be obtained by using first-order quench perturbation theory \cite{PQ1,PQ2} as we showed in \cite{IsingQuench}.}. An extraordinary feature of this formula is that it is obtained from a form factor expansion of the l.h.s. In other words, the emergence of an exponential on the r.h.s. is the result of exact resummation of infinitely many contributions. More details on the nature of this expansion are given in the next sections.}  As we will show, also the SREE displays linear growth, in fact, for this quench, the growth rate is identical to that of the total R\'enyi entropies, and oscillatory corrections which are distinct from those found for the total entropy. In a recent work \cite{Salvo_2025} it was argued that, crucially, the relaxation dynamics depends on the locality properties of the observable under consideration. In particular, it was shown that for local operators, additional power-law decaying terms are present. This implies that the post-quench dynamics of symmetry fields, such as the order field $\mu$ and twist fields $\TT, \TT_\mu$, is distinct from that of local operators. In particular, for twist fields additional polynomial tails are absent. Another observation investigated in \cite{pompon} is that the dynamics of the one-point functions of different symmetry fields in the same model (such as $\mu$ and $\TT$ here) is driven by exponential decay, with decay rates that are simply related to each other. 

\medskip

This paper is organised as follows: in Section \ref{2} we characterise the quench protocol and introduce the form factor approach. In Section \ref{3} we present the main calculations, which follow very closely the structure of \cite{IsingQuench}. In Section \ref{4} we write the new formula for the one-point function of the composite branch point twist field  and discuss the differences and similarities with existing formulae for the branch point twist field and the order parameter. We then show how the SREE can be obtained from these formulae and discuss its main properties. We conclude in Section \ref{5}.

\section{Quench Protocol and Form Factor Approach}
\label{2}
The main technique that is employed in \cite{SE,IsingQuench} is based on properties of the form factors of $\sigma$ and $\TT$, as well as on a certain simple regularisation of the singularities known as $\kappa$-regularisation. Form factors, that is matrix elements of the fields between the ground state and an excited state, come into play because the mass quench of the Ising field theory can be described exactly through a relationship between the pre- and post-quench ground state. Calling $|\Omega\ket$ the non-normalised initial state $|\Omega\rangle:=\sqrt{\langle\Omega|\Omega\rangle}|0\rangle$, with $|0\rangle$ the ground state of the pre-quench Hamiltonian (with mass scale $m_0$), this can be expressed in terms of eigenstates of the post-quench Hamiltonian in a form which is reminiscent of boundary states \cite{Gosh}
 \beq
 |\Omega\ket=\exp\left[\int_0^\infty \frac{d\theta}{2\pi} K(\theta) a^\dagger(-\theta) a^\dagger(\theta) \right]|\tilde{0}\ket\,.
 \label{psio}
 \eeq 
Here $|\tilde{0}\ket$ is the vacuum of the post quench Ising field theory (with mass gap $m$); $a^\dagger(\theta)$ is the fermionic creation operator,  and $K(\theta)$ is the function 
\begin{equation}
K(\theta)=i\tan\left[\frac{1}{2}\tan^{-1}(\sinh\theta)-\frac{1}{2}\tan^{-1}\left(\frac{m}{m_0}\sinh\theta\right)\right]:=i\hat{K}(\theta)\,,
\label{KK}
\end{equation}
which can be found in \cite{SFM,MFi}.
The integral in~\eqref{psio} is over the rapidity $\theta$ which, as usual, parametrises the energy $e$ and momentum $p$ of the one-particle  state $|\theta\rangle:= a^{\dagger}(\theta)|\tilde{0}\rangle$ as $e(\theta)=m\cosh\theta$ and $p(\theta)=m\sinh\theta$. The normalisation of the one-particle states is $\langle\theta|\theta'\rangle=2\pi\delta(\theta-\theta')$ and similarly for higher-particle states. 

Quenches of the type (\ref{psio}) where studied in detail in \cite{MFi,SFM}. A derivation of the function (\ref{KK}) is given in Appendix A of \cite{SFM}. These states have the same structure of the boundary states first described by Ghoshal and Zamolodchikov \cite{Gosh}. Their structure neatly fits with the quasiparticle picture put forward in \cite{EEquench,quench,quench2} as the initial state (\ref{psio}) can be regarded as a coherent superposition of quasiparticle pairs, also known as a squeezed coherent state. Exact solvability of the quench dynamics, which is generally not possible, has been also related~\cite{Piroli2017} to initial states analogous to~\eqref{psio}, see for instance \cite{PQ1}. In the $n$-copy theory, the above simply generalises to
\beq
|\Omega\ket_n=\exp\left[\sum_{j=1}^n\int_0^\infty \frac{d\theta}{2\pi} K(\theta) a_j^\dagger(-\theta) a_j^\dagger(\theta) \right]|\tilde{0}\ket_n\,,
 \label{nstate}
 \eeq 
where $a_j^\dagger(\theta)$ is the fermionic creation operator in copy $j$. We denote by
$|\theta_1,\ldots, \theta_k\ket_{j_1,\ldots,j_k;n}$  an element of an orthonormal basis in the replicated ($in$ or $out$) Hilbert space  consisting of $k$ particles with rapidities $\theta_i$ and copy labels $j_i$, $i=1,\ldots, k$. The energy and momentum of  multi-particle states are  the sum of the energies and momenta of their one-particle constituents. 

In such a framework, the moments of the symmetry resolved entropy can be written as
 \begin{equation}
 Z_n(0):=\frac{ \varepsilon^{2\Delta_n} {}_n \bra \Omega| \TT(0,t)|\Omega\ket_n}{{}_n \bra \Omega|\Omega\ket_n}\quad \mathrm{and}\quad  Z_n(1):=\frac{ \varepsilon^{2\Delta^\mu_n} {}_n \bra \Omega| \TT_\mu(0,t)|\Omega\ket_n}{{}_n \bra \Omega|\Omega\ket_n}\,,
 \label{t_SE}
 \end{equation}
 and the symmetry resolved partition functions $\mathcal{Z}_n(0)$, $\mathcal{Z}_n(1)$ are the discrete Fourier transforms of these quantities, with $\Delta_n$ and $\Delta_n^\mu$ the (known) conformal dimensions of the fields \cite{orbifold,CDL,GS}
 \beq 
\Delta_n=\frac{1}{48}\left(n-\frac{1}{n}\right)\,, \qquad \Delta_n^\mu=\frac{1}{48}\left(n-\frac{1}{n}\right) + \frac{1}{16 n}\,.
 \eeq 
 While $Z_n(0)$ is precisely the quantity computed in \cite{IsingQuench}, $Z_n(1)$ is computed in this paper.  Since the fields $\TT$ and $\TT_\mu$ satisfy the same exchange relations with the fermions and their matrix elements satisfy the same form factor equations as well, the computation of $Z_n(1)$ is very similar to that already performed for $Z_n(0)$. While this makes it easier to obtain analytical results, it will still provide us with the first analytic computation based on branch point twist fields for the SREE out of equilibrium and allow us to establish not only its linear growth but also the nature of subleading corrections for large times. 
\subsection{Form Factor Expansion}
 Substituting the representation \eqref{nstate} of the replicated initial state into~\eqref{t_SE}, both numerator and denominator admit a formal expansion as sums of integrals of matrix elements in the post quench basis. Proceeding as in~\cite{SE}, we write 
 \beq
  {}_n \bra \Omega| \TT_\mu(0,t)|\Omega\ket_n:=\tilde{\tau}_n^\mu\sum_{k_1, k_2=0}^\infty  C_{2k_1, 2k_2}(t)\,,
  \label{1po}
 \eeq
 where $\tilde{\tau}_n^\mu:={}_n\bra \tilde{0}|\TT_\mu(0,0)|\tilde{0}\ket_n$ is the vacuum expectation value in the post-quench ground state and 
 \begin{align}
&\tilde{\tau}_n^\mu C_{2k_1, 2k_2}(t)= \frac{1}{k_1! k_2!} \sum_{j_1,\ldots, j_{k_1}=1}^n  \sum_{p_1,\ldots, p_{k_2}=1}^n  \nonumber\\
& \times \left[\prod_{s=1}^{k_1 }\int_{0}^\infty \frac{d\theta'_{s}}{2\pi}K(\theta'_{s})^* e^{2 i t E(\theta'_{s})}\right] \left[\prod_{r=1}^{k_2 }\int_{0}^\infty \frac{d\theta_{r}}{2\pi}K(\theta_{r})e^{-2  i t E(\theta_r)}\right]\nonumber\\
 & \times  {}_{n;j_1j_1\ldots j_{k_1} j_{k_1}} \bra \theta'_1,-\theta'_1, \ldots, \theta'_{k_1},-\theta'_{k_1}|\TT_\mu(0,0)| -\theta_{k_2},\theta_{k_2},\dots, -\theta_{1}, \theta_1 \ket_{p_{k_2}p_{k_2}\ldots p_1 p_1;n}\,.
  \label{C22}
 \end{align}
 \fed{This formula results from expanding the state $|\Omega\ket_n$, starting from the exponential \eqref{nstate} term by term. Since the exponential consists of quasi-particle pairs of opposite rapidities, the one-point function becomes a sum of averages over states containing an even number of excitations which are organised in pairs of opposite momenta (rapidity).} Analogously
 \beq
 {}_n \bra \Omega|\Omega\ket_n:=\sum_{k=0}^\infty  Z_{2k}\,,
 \label{zpo}
 \eeq
with
  \beqa
&& Z_{2k}= \frac{1}{(k!)^2} \sum_{j_1,\ldots, j_{k}=1}^n  \sum_{p_1,\ldots, p_{k}=1}^n\left[\prod_{s=1}^{k }\int_{0}^\infty \frac{d\theta'_{s} d\theta_s}{(2\pi)^2}K(\theta'_{s})^* K(\theta_{s})\right]\nonumber\\
 && \times  {}_{j_1j_1\ldots j_{k} j_{k}} \bra \theta'_1,-\theta'_1, \ldots, \theta'_{k},-\theta'_{k}| -\theta_{k},\theta_{k},\dots, -\theta_{1}, \theta_1 \ket_{p_{k}p_{k}\ldots p_1 p_1} \quad \mathrm{for}\quad  k>0\,,
 \label{Z22}
 \eeqa
 and $Z_0=1$. \fed{Putting together \eqref{C22} and \eqref{Z22} we can formally expand the ratio \eqref{t_SE} in powers of the function $K$ as}
 \begin{equation}
 \label{d_series}
 \frac{{}_n \bra \Omega| \TT_\mu(0,t)|\Omega\ket_n}{{}_n \bra \Omega|\Omega\ket_n}:=\tilde{\tau}_n^\mu\sum_{k_1,k_2=0}^{\infty} D_{2k_1 2k_2}(t)\,,
 \end{equation}
with 
\beq 
D_{2k_1,2k_2}(t)=\sum_{p=0}^{\min(k_1,k_2)} \tilde{Z}_{2p} C_{2(k_1-p),2(k_2-p)}(t)\,, \label{eq:D_2k_2l_original}
\eeq 
where $\tilde{Z}_{2p}$ are the expansion coefficients of the inverse of the norm, i.e. $\sum_{k,p=0}^\infty Z_{2k}\tilde{Z}_{2p}=1$. \fed{A remarkable feature of this resummation is that divergencies arising due to coinciding rapidities in \eqref{C22} and \eqref{Z22} are cancelled out in (\ref{d_series}). A good presentation of how these cancellations occur can be found for instance in \cite{Leclair:1996bf}.} In \cite{SE,IsingQuench} a calculation of all terms up to order $O(K^2)$ was performed and the same can be done for the field $\TT_\mu$.  A useful observation is that the operator $\TT_\mu$ reduces to $\mu$ for $n=1$, which means that we should recover the results of \cite{SE} for the one-point function of magnetisation if we set $n=1$. This provides a benchmark for our results. 
 
 The matrix elements of the composite twist field in~\eqref{C22}  can be related to the so-called elementary form factors~\cite{KW,smirnovbook,entropy,CTFMichele}.  
 The transformation that relates the two functions  is called crossing. Consider for instance the matrix element $_{n;j_1}\bra \theta_1|\mathcal{T}_\mu(0,0)|\theta_2\ket_{j_2;n}$. This can be written as
 \beqa
\label{first}
	_{n;j_1}\bra \theta_1|\mathcal{T}_\mu(0,0)|\theta_2\ket_{j_2;n} &=&\tilde{\tau}_n^\mu  {\,} _{n;j_1}\bra \theta_1|\theta_2\ket_{j_2;n} 
    +{}_n\bra \tilde{0}|\mathcal{T}_\mu(0,0)|\theta_1+i\pi-i\eta,\theta_2\ket_{j_1,j_2;n}\nonumber\\
    &=&2\pi\,\tilde{\tau}_n^\mu \, \delta(\theta_{12})\delta_{j_1j_2}+F_2^{j_1 j_2}(\theta_{12}+i\pi-i\eta)\,,
\eeqa
where $\theta_{12}:=\theta_1-\theta_2$, $\eta$ is a small positive parameter and $F_{2}^{j_1 j_2}(\theta)$ is the two-particle form factor of the operator $\TT_\mu$ in copies $j_1, j_2$. 
 The form factors of the composite twist field $\TT_\mu$ where first obtained in \cite{SymResFF} and have the Pfaffian structure typical of the Ising model, that is 
\beq
F_{2k}^{11\ldots 1}(\theta_1,\ldots,\theta_{2k};n):={}_n\bra \tilde{0}|\TT_\mu(0,0)|\theta_1,\ldots,\theta_{2k}\ket_{1,\ldots, 1;n} = \tilde{\tau}_n^\mu \, {\rm Pf}(W)\,,
\eeq 
with
\beq
\label{wfun}
W_{ij}:=w(\theta_{ij})=\frac{\sin\frac{\pi}{n}}{2n \sinh\left(\frac{i\pi - \theta_{ij}}{2n}\right)\sinh\left(\frac{i\pi + \theta_{ij}}{2n}\right)}\frac{\sinh\frac{\theta_{ij}}{n}}{\sinh\frac{i\pi}{n}}\,.
\eeq
 This formula differs from the form factors of $\TT$ only because $n$ is replaced by $n/2$ in the minimal part of the form factor (\textit{i.e.} the ratio of sinh functions on the r.h.s.). However, this small change leads to some important differences, the main one being the asymptotic properties
\beq
 \lim_{\theta \rightarrow \pm \infty} w(\theta)=\pm \frac{i}{n} \qquad \mathrm{and} \qquad \lim_{n\rightarrow 1} w(\theta)=i \tanh\frac{\theta}{2}\,.
\label{intprop}
\eeq 
Note that the last equality simply shows that the two-particle form factor of $\TT_\mu$ reduces to that of $\mu$ for $n=1$, as expected \cite{YZam} \fed{in the disordered phase}. This extends to higher-particle form factors too, although in practice most of the work we need to do will involve only up to four particle form factors (since these are the ones involved in terms of order $K^2$). 

The relation (\ref{first}) generalises to matrix elements involving states with a larger number of particles~\cite{smirnovbook}. The shift by $i\eta$ makes the function $F_{2}^{j_1 j_2}(\theta)$ on the right hand side of \eqref{first} regular for $\theta\rightarrow i\pi$.  There are however additional sources of divergences related to the normalisation of the asymptotic states in infinite volume, see the $\delta$-function in \eqref{first}.

These infinite volume singularities are cancelled by similar singularities in the denominator in \eqref{Z22} in the combination \eqref{eq:D_2k_2l_original}. The precise way in which this cancellation occurs has been extensively investigated and is well understood by considering the theory in finite volume $V$ and using the volume as a regulator \cite{PT1,PT2}. 

Sometimes, like in the present case, a simpler regularisation scheme can also be employed. In~\cite{SE} so-called $\kappa$-regularisation \cite{EK1,EK2} was used.  The idea is simply to shift coinciding rapidities by a real value $\kappa$  (or several values $\kappa_i$ for multi-particle states) so that the singularities are avoided and then take the limit $\kappa \rightarrow 0$ in a controlled way. 
See~\cite{SE,IsingQuench} for applications.

\section{Computation of \texorpdfstring{$Z_n(1)$}{Zn(1)}}
\label{3}
We now come to the main section of our paper, that is the new computation of the charged moment $Z_n(1)$, the normalised one-point function of the field $\TT_\mu$. As in \cite{IsingQuench}, the computation is organised around terms of order $K$ and $K^2$, and the computation steps are identical except for the subtleties due to the distinct nature of the form factors of $\TT_\mu$.
\subsection{Contributions at \texorpdfstring{$O(K)$}{O(K)}}

Apart from  a trivial $K$-independent term, corresponding to $C_{00}=D_{00}=1$ in~\eqref{1po}, the leading term in the $K$ expansion of~\eqref{t_SE} is $O(K)$ and is given in terms of the two-particle form factors of the field. Using the notations introduced earlier (recall the definition of the hatted function $\hat{K}(\theta)$ in (\ref{KK})), we have 
 \beqa
C_{2,0}(t)+C_{0,2}(t)&=& n\left[ \int_0^\infty \frac{d\theta}{2\pi} K(\theta)^* w(2\theta) e^{2itE(\theta)} +\int_0^\infty \frac{d\theta}{2\pi} K(\theta) w(2\theta)^* e^{-2itE(\theta)}\right] \nonumber\\
&=&
-  \int_0^\infty \frac{d\theta}{2\pi}  \hat{K}(\theta) \frac{\sinh\frac{2\theta}{n}}{\sinh\frac{i\pi-2\theta}{2n} \sinh\frac{i\pi+2\theta}{2n}} \cos\left[2m t\cosh\theta \right]\,,
\label{C2002}
\eeqa
where we have used~\eqref{wfun}. Notice that the expansion of the denominator in~\eqref{zpo} starts as $1+O(K^2)$, therefore, see~\eqref{eq:D_2k_2l_original}, $C_{2,0}+C_{0,2}=D_{2,0}+D_{0,2}$.
 At large times, according to stationary phase analysis, we can expand the integrand in~\eqref{C2002} close to $\theta=0$ and observe that 

\beq
\hat{K}(\theta)= \frac{\alpha}{2} \theta+O(\theta^3)\,,
\eeq
with $\alpha$ defined as
\beq
\label{alpha}
\alpha:= 1-\frac{m}{m_0}=-\frac{\delta m}{m_0}\,, \quad \mathrm{with}\quad \delta m:=m-m_0~.
\eeq
 By retaining only contributions up to  $O(K)$,  the one-point function of the composite twist field for $mt\gg 1$ is given by
\begin{equation}
\frac{{}_n\bra \Omega |\TT_\mu(0,t)|\Omega\ket_n}{{}_n\bra \Omega|\Omega\ket_n}=\tilde{\tau}_n^\mu \left(1-\frac{\alpha}{8 \sqrt{\pi } n \sin^2\frac{\pi }{2 n}}  \frac{\cos \left(2 m t-\frac{\pi}{4}\right)}{(mt)^{3/2}}+\dots\,\right)+O(K^2)\,.
\label{mismo2}
\end{equation}
For $|\alpha|\ll 1$   the same result can be derived from a perturbation theory approach~\cite{PQ1}. The result is very similar to that obtained for the standard twist field $\TT$. In particular, we have the same type of oscillatory function and the same power-law suppression $(mt)^{-3/2}$ of the oscillations for large times.  For $n=1$ we recover the expansion for the field $\mu$ as given in \cite{SE}.

\subsection{Contribution at Order \texorpdfstring{$O(K^2)$}{O(K2)}}

The next corrections are at the $O(K^2)$ order, where we have the contributions $C_{2,2}(t)$, $C_{0,4}(t)$ and {\color{black} $C_{4,0}(t)$}. We first concentrate on the term $C_{2,2}(t)$, which takes the form
\begin{equation}\begin{split}\label{eq:C22}
    \tilde{\tau}_n^\mu C_{2,2}(t) =& n \sum_{j = \fed{1}}^{n} \int_{0}^{+\infty} \frac{\dd \theta' \dd\theta}{(2\pi)^2}\, K(\theta')^* K(\theta) e^{2\ii m t \left( \cosh\theta' - \cosh\theta \right)}\, {}_{11}\bra \theta', -\theta'| \mathcal{T}_\mu(0,0) | -\theta +\kappa, \theta+\kappa \ket_{jj}\\
    =& n \sum_{j = \fed{1}}^{n} \int_{0}^{+\infty} \frac{\dd \theta' \dd\theta}{(2\pi)^2}\, M(\theta', \theta; t)\, {}_{11}\bra \theta', -\theta'| \mathcal{T}_\mu(0,0) | -\theta +\kappa, \theta+\kappa \ket_{jj},
\end{split}
\end{equation}
where we introduced the function
\begin{equation}
    M(\theta_1, \theta_2; t) = \widehat{K}(\theta_1) \widehat{K}(\theta_2) e^{2 \ii m t \left[ \cosh \theta_1 - \cosh \theta_2 \right] }.
\end{equation}
 The computation of the integral in equation \eqref{eq:C22} requires the use of crossing and exchange symmetries of the form factors. First, we rewrite the matrix element inside the integral using the crossing relation
\begin{equation}\begin{split}
\label{crossing equation}
    &{}_{11}\bra \theta', -\theta'| \TT_\mu(0,0) | -\theta +\kappa, \theta+\kappa \ket_{jj} =\\
    =& \left ( 2\pi \right )^2 \tilde{\tau}_n^\mu  \left [ \delta(\theta'-\theta-\kappa) \delta(-\theta'+\theta-\kappa) \delta_{1j} - \delta(\theta'+\theta-\kappa) \delta(-\theta'-\theta-\kappa) \delta_{1j} \right ] +\\
    &+ 2\pi \, \big [ \delta(\theta'-\theta-\kappa)\delta_{1j} F_{2}^{1j}\!\left ( -\theta' + \ii\pi - \ii\eta + \theta - \kappa \right ) + \delta(-\theta'+\theta-\kappa)\delta_{1j} F_{2}^{1j}\!\left ( \theta' + \ii\pi - \ii\eta - \theta - \kappa \right ) +\\
    &\hspace{.7cm} - \delta(\theta'+\theta-\kappa)\delta_{1j} F_{2}^{1j}\!\left ( -\theta' + \ii\pi - \ii\eta - \theta - \kappa \right ) - \delta(-\theta'-\theta-\kappa)\delta_{1j} F_{2}^{1j}\!\left ( \theta' + \ii\pi - \ii\eta + \theta - \kappa \right ) \big] +\\
    &+ F_{4}^{11jj}\!\left(\theta'+\ii\pi-\ii\eta_1, -\theta'+\ii\pi-\ii\eta_2, -\theta+\kappa, \theta+\kappa \right).
\end{split}\end{equation}
Separating the terms containing respectively two, one and no delta functions, we cast the contribution $C_{2,2}(t)$ as the sum of three terms
\begin{equation}
    C_{2,2}(t) = C_{2,2}^{(0)}(t) + C_{2,2}^{(2)}(t) + C_{2,2}^{(4)}(t).
    \label{contributionsC22}
\end{equation}
Let us first compute the contribution $C_{2,2}^{(2)}(t)$. We extend the domain of integration of $\theta$ and $\theta'$ from $-\infty$ to $\infty$, obtaining
\begin{equation}\begin{split}
    \tilde{\tau}_n^\mu C_{2,2}^{(2)}(t) &= n \sum_{j=1}^{n} \int_{-\infty}^{+\infty}\frac{\dd \theta \dd \theta'}{(2\pi)^2}\, M(\theta', \theta; t)\,  F^{1j}_{2}(-\theta'+\theta+\ii \pi-\ii\eta-\kappa) \left [ (2\pi) \delta(\theta'-\theta-\kappa)\delta_{1j} \right ]\\
    &= n\, \tilde{\tau}_n^\mu  w(\ii\pi -2\kappa - \ii\eta) \int_{-\infty}^{+\infty}\frac{\dd \theta }{2\pi}\, M(\theta+\kappa,\theta;t).
    \label{C222}
\end{split}\end{equation}
Since the delta function imposes that the incoming and the outgoing rapidities are identical, now the form factor does not depend on $\theta$ and is computed in the vicinity of the kinematic pole $\ii \pi$.
This gives a divergent contribution that is regularised by the small rapidity shift $\kappa$. 
Expanding the expression around the pole $2\kappa = -\ii\eta$, we obtain
\begin{equation}\label{eq:C22(2)}\begin{split}
    C_{2,2}^{(2)}(t) &\approx \left [- \frac{\ii n}{2\fed{\kappa} + \ii \eta} + \frac{\cos \frac{\pi}{n}}{2 \sin \frac{\pi}{n}} + O(\kappa) \right ] \\
    & \times \int_{-\infty}^{+\infty}\frac{\dd \theta}{2\pi} \left [ \widehat{K}(\theta)^2 + \kappa \widehat{K}(\theta) \frac{\dd \widehat{K}(\theta)}{\dd \theta} + 2\ii mt \kappa \widehat{K}(\theta)^2 \sinh \theta + O(\kappa^2)  \right ]\\
    &\approx - \frac{\ii n}{2\fed{\kappa} + \ii \eta} \int_{-\infty}^{+\infty}\frac{\dd \theta}{2\pi} \widehat{K}(\theta)^2 + \frac{\cos \frac{\pi}{n}}{2 \sin \frac{\pi}{n}} \int_{-\infty}^{+\infty}\frac{\dd \theta}{2\pi} \widehat{K}(\theta)^2 + O(\kappa) \\
    &= - \frac{\ii n}{2\fed{\kappa} + \ii \eta} \int_{-\infty}^{+\infty}\frac{\dd \theta}{2\pi} \widehat{K}(\theta)^2 + A \cos \frac{\pi}{n}\,,
\end{split}
\end{equation}
where the terms proportional to the derivative of $\widehat{K}(\theta)$ and to $\sinh \theta$ vanish identically because of symmetry and
\begin{equation}
\label{A}
    A = \frac{1}{2 \sin \frac{\pi}{n}} \int_{-\infty}^{+\infty}\frac{\dd \theta}{2\pi} \widehat{K}(\theta)^2\,
\end{equation}
is the same constant found in the expansion for the standard branch point twist field and shown to provide a correction to the vacuum expectation value of the field\footnote{It is interesting to observe that there is no constant correction to the vacuum expectation value of the operator $\mu$ and indeed none was computed in \cite{SE}. This gives an indication of the subtle nature of the interplay between the limit $n\rightarrow 1$ and the limit $\kappa\rightarrow -i\eta/2$. The two limits do not commute. If we do the expansion \eqref{eq:C22(2)} first, and then try to set $n=1$ we find that $A$ is singular. On the other hand, if we set $n=1$ from the start in \eqref{C222} we find that $A=0$.}.
We see that in the limit $\kappa\to 0, \eta \to 0$, the contribution $C_{2,2}^{(2)}(t)$ in \eqref{eq:C22(2)} is comprised of a constant $A \cos \frac{\pi}{n}$ and a divergent term which will be cancelled off by an opposite term in $C_{2,2}^{(4)}(t)$. 

Next we consider the contribution $C_{2,2}^{(4)}(t)$.
Using the a Pfaffian structure of the four particle form factor
\begin{equation}\label{eq:F4}\begin{split}
    &F_4^{11jj}\!\left( \theta'+ \ii \pi - \ii \eta_1, - \theta'+\ii\pi -\ii\eta_2, -\theta+\kappa, \theta+\kappa \right) =\\
    =&\, F_2^{11}\!\left ( 2\theta' - \ii\eta_{12} \right ) F_2^{jj}\!\left ( 2\theta' \right ) - F_2^{1j}\!\left ( \theta' + \theta + \ii\pi - \ii\eta_{1} -\kappa \right ) F_2^{1j}\!\left ( - \theta' - \theta + \ii\pi - \ii\eta_{2} - \kappa \right ) + \\
    &\hspace{4.2cm} + F_2^{1j}\!\left ( \theta' - \theta + \ii\pi - \ii\eta_{1} -\kappa \right ) F_2^{1j}\!\left ( - \theta' + \theta + \ii\pi - \ii\eta_{2} - \kappa \right ),
\end{split}\end{equation}
we see that we can separate the term $C_{2,2}^{(4)}$ into a \lq\lq connected\rq\rq part $I(t)$ and a disconnected or factorised part $I'(t)$ (with respect to the $\theta$, $\theta'$ variables)
\begin{equation}\begin{split}
    C_{2,2}^{(4)}(t) &= n \sum_{j=1}^{n} \int_{0}^{+\infty}\frac{\dd \theta \dd \theta'}{(2\pi)^2}\, M(\theta', \theta; t) F_{4}^{11jj}(\theta'+ \ii \pi - \ii \eta_1, - \theta'+\ii\pi -\ii\eta_2, -\theta+\kappa, \theta+\kappa)\\
    &= I(t) + I'(t).
\end{split}\end{equation}
The disconnected term $I'(t)$ factorises into the product of $C_{2,0}(t)$ and $C_{0,2}(t)$
\begin{equation}
    I'(t) = n^2 \int_0^{+ \infty} \frac{\dd \theta \dd \theta'}{(2\pi)^2}\,  M(\theta', \theta; t)\, w(2\theta') w(-2\theta) = C_{2,0}(t)\, C_{0,2}(t) = \left | C_{2,0}(t) \right |^2.
\end{equation}
{\fed{The contribution $I'(t)$ together with the contribution $C_{2,2}^{(0)}=\frac{1}{2}(C_{2,0}(t)^2+C_{0,2}(t)^2)$ add up to $\frac{1}{2}(C_{2,0}(t)+C_{0,2}(t))^2$. This suggests that the $O(K)$ terms found earlier $C_{2,0}(t)+C_{0,2}(t)$ are just the beginning of the expansion of the exponential $e^{C_{2,0}(t)+C_{0,2}(t)}$. Indeed, the fact that this guess is correct and exact at all orders was shown in \cite{IsingQuench}}.
As for the disconnected part, 
noting that $M(\theta', \theta; t)=-M(-\theta',\theta;t)=-M(\theta',-\theta;t)$ we can extend the domain of integration of $I(t)$ to $\theta$ from $-\infty$ to $\infty$. 
The sum over the replicas in the connected part $I(t)$ can be carried out by using the summation formula for composite twist field form factors \cite{Horvath:2020vzs,Horvath:2021fks,CTFMichele}
\begin{equation}\label{eq:sumcomposite}
    \sum_{j = 0}^{n-1} w\!\left(-x+2\pi \ii j\right ) w\!\left( y+2\pi\ii j \right ) = \frac{- \ii\sinh\!\frac{x+y}{2}}{2 \cosh\!\frac{x}{2} \cosh\!\frac{y}{2}} \left [ w\!\left ( x+y+\ii \pi \right ) +  w\!\left ( x+y-\ii \pi \right ) \right ] -\frac{1}{n}\,,
\end{equation}
obtaining
\begin{eqnarray}
    I(t) &=& n \int_{0}^{+\infty} \frac{\dd \theta}{2\pi} \int_{-\infty}^{+\infty}\frac{ \dd \theta'}{2\pi}\, M(\theta', \theta; t) \sum_{j= 0}^{n-1} w(-\theta'+\theta-\ii\pi+\ii\eta_1+2\pi\ii j) w(\theta'-\theta-\ii\pi+\ii\eta_2+2\pi\ii j)\nonumber\\
    &=& n \int_{0}^{+\infty} \frac{\dd \theta}{2\pi} \int_{-\infty}^{+\infty}\frac{ \dd \theta'}{2\pi}\, M(\theta', \theta; t)\,\frac{H_{\mu}(\theta', \theta)}{2 \sinh\!\left ( \frac{\theta'-\theta-\kappa -\ii\eta_1}{2}\right) \sinh\!\left ( \frac{\theta'-\theta+\kappa +\ii\eta_2}{2}\right )}\,,
\label{eq:I(t)}
\end{eqnarray}
with
\beq 
H_\mu(\theta',\theta):= -\ii \sinh\left(\theta'-\theta-\frac{i\eta_{12}}{2}\right) [w(2\theta'-2\theta-\ii \eta_{12} +\ii \pi)+w(2\theta'-2\theta-\ii \eta_{12} -\ii \pi)]\,.
\label{Hmu}
\eeq 
Notice that the contribution resulting from the term $-\frac{1}{n}$ term in \eqref{eq:sumcomposite} cancels out exactly because of the parity of $M(\theta',\theta;t)$. 
The integrand in \eqref{eq:I(t)} has poles at $\theta' = \theta +\kappa+\ii\eta_1$ and $\theta' = \theta -\kappa-\ii\eta_2$.
In order to compute the leading asymptotic behaviour of \eqref{eq:I(t)}, we deform the contour of integration to $\mathcal{C} = \mathcal{C}_1 \cup \mathcal{C}_2 \cup \mathcal{C}_3$, with
\begin{align}
    \mathcal{C}_1 &= \left \{ x - s + \ii \phi\, \big |\, x \in \left [ -\infty, 0 \right ] \right \} ,\\
    \mathcal{C}_2 &= \left \{ -s +\ii x\, \big |\, x \in \left [-\phi, \phi \right ] \right \} ,\\
    \mathcal{C}_3 &= \left \{ x - s + \ii \phi\, \big |\, x \in \left [ 0, +\infty\right ] \right \} .
\end{align} 
When deforming the integral, we pick up the residue of the kinematic pole at $\theta' = \theta- \kappa - \ii \eta_1$ with negative  orientation. Multiplying the residue by $-2\pi \ii$ yields 
\beq 
n \int_{0}^{\infty} \frac{\dd\theta'}{2\pi}\, M(\theta',\theta'-\kappa;t) \left [ w(2\kappa - \ii\eta + \ii\pi) + w(2\kappa - \ii\eta - \ii\pi) \right ]\,, 
\eeq
which, when expanding around $\kappa=0$, leads to
\begin{equation}\begin{split}
    &\approx n \left[ \frac{\ii}{\kappa + \frac{\ii\eta}{2}} + O(\kappa) \right] \int_{0}^{+\infty} \frac{\dd \theta'}{2\pi} \left [\widehat{K}(\theta')^2 - \kappa \widehat{K}(\theta') \frac{\dd \widehat{K}(\theta')}{\dd \theta'} + 2\ii mt \kappa \widehat{K}(\theta')^2 \sinh \theta'  + O(\kappa^2) \right ]\\
    &\approx \frac{\ii n}{2\fed{\kappa} + \ii \eta} \int_{-\infty}^{+\infty}\frac{\dd \theta}{2\pi} \widehat{K}(\theta)^2 - 2 n m t\int_{0}^{+\infty} \frac{\dd\theta}{2\pi}\, \widehat{K}(\theta)^2 \sinh \theta\\
    &= \frac{\ii n}{2\fed{\kappa} + \ii \eta} \int_{-\infty}^{+\infty}\frac{\dd \theta}{2\pi} \widehat{K}(\theta)^2 - \frac{n\Gamma}{2} mt\,.
\end{split}\end{equation}
We see that the residue of the kinematic pole contains a divergent part which is cancelled by the opposite contribution in \eqref{eq:C22(2)} and a contribution linear in time, which provides the leading term at large $mt$ in the linear growth predicted by the quasiparticle picture, the growth rate being $\frac{n \Gamma}{2}$ with
\begin{equation}
    \Gamma = \frac{2}{\pi} \int_{0}^{+\infty} \dd\theta\, \widehat{K}(\theta)^2 \sinh \theta\, .
    \label{G}
\end{equation}
This is exactly the same growth rate as found for the total entropy in \cite{IsingQuench}.
The remaining contribution to the integral (\ref{eq:I(t)}), which
is now regular and asymptotically decreasing in $t$, is
\begin{equation}
    R(t) =  n \int_{0}^{+\infty} \frac{\dd \theta}{2\pi} \int_{\mathcal{C}}\frac{ \dd \theta'}{2\pi}\, M(\theta', \theta; t)\,\frac{H_{\mu}(\theta', \theta)}{2 \sinh\!\left ( \frac{\theta'-\theta-\kappa -\ii\eta_1}{2}\right) \sinh\!\left ( \frac{\theta'-\theta+\kappa +\ii\eta_2}{2}\right )}\,.
\end{equation}
A convenient way to compute the integral is to differentiate it twice with respect to $t$.
This differentiation eliminates the poles and allows us to deform back to the original contour. 
We obtain
\begin{equation}
    \frac{\dd^2 R(t)}{\dd t^2} = - 4 m^2 n \int_{0}^{+\infty} \frac{\dd \theta'}{2 \pi} \int_{-\infty}^{+\infty} \frac{\dd \theta}{2 \pi} M(\theta', \theta; t) \, \frac{H_{\mu}(\theta', \theta) \left( \cosh \theta' - \cosh \theta \right )^2}{2 \sinh^2\left(\frac{\theta'-\theta}{2}\right)}\,,
\end{equation}
which can be computed in the stationary phase approximation. 
In this approximation, the leading contribution comes from $H_{\mu}(\theta', \theta) \approx 1 + O(\theta)$, yielding
\begin{equation}
    \frac{\dd^2 R(t)}{\dd t^2} \approx - 2 m^2 n \int_{-\infty}^{+\infty} \frac{\dd \theta' \dd \theta}{(2 \pi)^2}   \, \frac{\alpha^2}{4} \theta'^2 \theta^2 e^{\ii m t \left ( \theta'^2 - \theta^2 \right )} + O(t^{-5}) = - \frac{n \alpha^2}{32\pi mt^3} +O(t^{-5})\,.
\end{equation}
We can now integrate back, using the knowledge that the original $R(t)$ is decreasing with $t$ to eliminate any contributions linear in $t$ or constant, obtaining
\begin{equation}
    R(t) = - \frac{n\alpha^2}{64 \pi m t} + O(t^{-3})\,.
    \label{R}
\end{equation}
Again, we find that at this order the term $R(t)$ is identical to the correction found for $Z_n(0)$ in \cite{IsingQuench}.
The integrals are however different and we would see differences if we considered higher order corrections (higher powers of $(mt)^{-1}$). In summary, we have that 
\beq 
I(t)=-\frac{n \Gamma mt}{2}- \frac{n\alpha^2}{64 \pi m t} + O(t^{-3})\,,
\eeq 
a result which is identical to that obtained for the analogous terms in $Z_n(0)$.

\subsubsection{\texorpdfstring{$C_{0,4}(t)$}{C04} and \texorpdfstring{$C_{4,0}(t)$}{C40} terms}

The remaining terms at the $O(K^2)$ order are those with four incoming (outgoing) and no outgoing (incoming) particles, that is, $C_{4,0}(t)$ ($C_{0,4}(t)$).
As we will see, these terms contribute to reproduce the exponential of the result at order $O(K^2)$.
Since the two terms are complex conjugate of each other, $C_{4,0}(t) = C_{0,4}(t)^*$, we focus on $C_{0,4}(t)$.
We write it as
\begin{equation}
    C_{0,4}(t) = -\frac{n}{2} \sum_{j = 1}^{n} \int_{0}^{+\infty} \frac{\dd\theta \dd\theta'}{(2\pi)^2}\,N(\theta, \theta'; t)\, F_4^{11jj}\!\left( -\theta, \theta, -\theta', \theta'\right),
\end{equation}
where we introduced the function 
\begin{equation}
    N(\theta, \theta'; t) = \widehat{K}(\theta) \widehat{K}(\theta')\, e^{-2\ii m t \left [\cosh \theta + \cosh \theta' \right ]}.
    \label{N}
\end{equation}
Using the Pfaffian structure in Eq.~\eqref{eq:F4}, we can again separate the integral into a connected and disconnected part.
The disconnected part is proportional to the square of the $C_{0,2}$ term
\begin{equation}
    -\frac{n^2}{2}\int_{0}^{+\infty} \frac{\dd\theta \dd\theta'}{(2\pi)^2}\,N(\theta, \theta'; t)\, F_2^{11}\!\left( -\theta, \theta\right) F_2^{11}\!\left( -\theta', \theta'\right) = \frac{1}{2} C^2_{0,2}(t)\,,
\end{equation}
with a similar contribution $\frac{1}{2} C^2_{2,0}(t)$ coming from $C_{4,0}(t)$.

 Extending the domain of integration of $\theta'$ from $-\infty$ to $+\infty$ we can write the connected part of the integral again in terms of the function $H_\mu(\theta',\theta)$ which we introduced in Eq.~\eqref{Hmu}:
\begin{equation}\begin{split}
    & \frac{n}{2} \int_{0}^{+\infty} \frac{\dd\theta}{2\pi}\int_{-\infty}^{+\infty}\frac{\dd\theta'}{2\pi}\,N(\theta, \theta'; t) \left[\sum_{j = 1}^{n} w\!\left( -\theta +\theta' - 2\pi\ii(j-1) \right)  w\!\left( \theta - \theta' - 2\pi\ii(j-1) \right) \right]\\
    &= \frac{n}{2} \int_{0}^{+\infty} \frac{\dd\theta}{2\pi}\int_{-\infty}^{+\infty}\frac{\dd\theta'}{2\pi}\,N(\theta, \theta'; t)\, \frac{H_\mu\!\left( \theta, \theta'\right)}{2\, \cosh^2\left(\frac{\theta-\theta'}{2}\right)}\,,
\end{split}\end{equation}
where we used \eqref{eq:sumcomposite} and the term proportional to $-\frac{1}{n}$ cancels exactly due to symmetry. In order to obtain the leading contribution for large $mt$ we must carry out a calculation which is similar to our computation of $R(t)$, except that in this case the integrand has no poles on the real line and so we can directly proceed to expanding around small $\theta$. This gives 
\beqa
&& \frac{n}{2} \int_{0}^{+\infty} \frac{\dd\theta}{2\pi}\int_{-\infty}^{+\infty}\frac{\dd\theta'}{2\pi}\,N(\theta, \theta'; t)\, \frac{H_\mu\!\left( \theta, \theta'\right)}{2\, \cosh^2\left(\frac{\theta-\theta'}{2}\right)} \nonumber\\
&& \approx \frac{n}{4} e^{-4i mt}\int_{0}^{+\infty} \frac{\dd\theta}{2\pi}\int_{-\infty}^{+\infty}\frac{\dd\theta'}{2\pi}\,\frac{\alpha^2}{4} \theta'^2\theta^2 e^{-i mt (\theta^2+\theta'^2)}=\frac{i n \alpha^2}{128 \pi} \frac{e^{-4i mt}}{(mt)^3}+O(t^{-5})\,.
\eeqa 
Combining this result with its complex conjugate contribution, coming from the connected part of $C_{4,0}(t)$ we obtain the final contribution
\beq 
\frac{n \alpha^2 \sin(4mt)}{64 \pi (mt)^3}+O(t^{-5})
\eeq 
Therefore, the connected part is oscillatory and of order $O(t^{-3})$, an order that we will neglect in our final result.

\section{One-Point Functions and Symmetry Resolved Entropies}
\label{4}
We now have all order $K$ and order $K^2$ contributions resulting from the expansion of the state $|\Omega\ket_n$. These contributions include up to order $\alpha^2$ in the quench strength and up to order $(mt)^{-3/2}$ in the dimensionless time variable. Putting all contributions together, we find 
\begin{equation}\begin{split}
    \frac{_n \bra{\Omega}| \mathcal{T}_\mu(0, t) |{\Omega}\ket_n}{_n \bra{\Omega}|{\Omega}\ket_n} & = \tilde{\tau}_n^\mu  \bigg [  1 + C_{2,0}(t) + C_{0,2}(t) + A \cos \frac{\pi}{n} - \frac{n\Gamma}{2}\, mt +\\
    &\hspace{2cm} + \frac{1}{2} \left ( C_{2,0}(t) + C_{0,2}(t) \right )^2 + R(t) + O(K^3) + O(t^{-3}) \bigg ]\,,
    \label{Tmu}
\end{split}\end{equation}
with $A$, $\Gamma$ and $R(t)$ given by \eqref{A}, \eqref{G} and \eqref{R} respectively. Except for the factor $\cos\frac{\pi}{n}$ which multiplies $A$, the formula is identical to that found for the field $\TT$ in \cite{IsingQuench}, up to the appropriate replacement of the cumulants $C_{2,0}(t)$ and $C_{0,2}(t)$. This similarity is due to the Pfaffian structure of both sets of form factors, which lends itself to various resummations. In particular, as we can already see in (\ref{Tmu}) we have what appears to be the first few terms in the expansion of an exponential. Indeed, in \cite{IsingQuench} it was shown that this is indeed the case for all the leading terms shown in \eqref{Tmu}. In other words, by employing identical resummation arguments as presented in \cite{IsingQuench} we can show that 
\beqa
    \frac{_n\bra{\Omega}| \mathcal{T}_\mu(0, t) |{\Omega}\ket_n}{_n\bra{\Omega}|{\Omega}\ket_n} &\approx& \tilde{\tau}_n^\mu  \exp\!\left(C^\mu_{2,0}(t) + C^\mu_{0,2}(t) + A \cos \frac{\pi}{n} - \frac{n\Gamma}{2}\, mt + R(t) )  \right)\nonumber\\
    &=& \tilde{\tau}_n'^\mu  \exp{\left(- \frac{n\Gamma}{2} mt - \frac{n\alpha^2}{64 \pi mt}-\frac{\alpha}{8 \sqrt{\pi } n \sin^2\frac{\pi }{2 n}}  \frac{\cos \left(2 m t-\frac{\pi}{4}\right)}{(mt)^{3/2}}\right)}\,,
    \label{formula}
\eeqa 
at orders $O((mt)^{-3})$ and $O(K^2)$, with $\tilde{\tau}_n'^\mu$ the corrected vacuum expectation value
\beq 
\tilde{\tau}_n'^\mu=\tilde{\tau}_n^\mu e^{A \cos\frac{\pi}{n}}\,.
\label{corr}
\eeq 
This can be compared with the formula for the branch point twist field obtained in \cite{IsingQuench} and given by equation (\ref{formula2}).

Interestingly, the leading large time behaviour of the magnetisation as well as of both twist fields is characterised by the same exponential decay rate $\Gamma$. Indeed, it has been suggested that this relationship between the decay rate of one-point functions of branch point twist fields and of order parameters may be universal \cite{pompon}. Our computation suggests that it is in fact shared by all the symmetry fields in the theory. 

Formula \eqref{formula} \fed{reproduces} the result of \cite{SE} for the one-point function of the magnetisation for $n=1$ \fed{if we expand the exponential and consider just the leading contribution from the oscillatory and other subleading terms}. There is just a slight disagreement in the term of order $(mt)^{-1}$ where we get an extra factor $1/2$ in our case. 

Note also that there is no correction to the vacuum expectation value of the operator $\mu$, as argued in the footnote after equation \eqref{eq:C22(2)}.
Indeed, it is important to highlight again that although $\kappa$ regularisation is an effective way to compute the one-point functions above for $n$ integer greater than 1, it does pose problems both for $\TT$ and $\TT_\mu$ as soon as we try to take the limit $n \rightarrow 1$. In the case of $\TT_\mu$ the limit gives the expectation value of magnetisation computed in \cite{SE}. However, there is a subtlety regarding the corrected vacuum expectation value $\tau_n'^\mu$. The correction (\ref{corr}) is divergent when $n\rightarrow  1$. However, we know from direct computation for $\mu$ that $A$ should go to zero when $n \rightarrow  1$, so $\tilde{\tau}_1'^\mu=\tilde{\tau}_1^\mu=\bra \tilde{0}|\mu(0,0)|\tilde{0} \ket$.

Issues are more serious for the field $\TT$. First of all, $\TT$ reduces to the identity field when $n=1$ and so the ratio (\ref{formula2}) should tend to 1 when $n\rightarrow 1$. This is however not the case here, even though it works for the oscillatory part of the exponential. Indeed, this oscillatory part is the only part that admits a well-defined analytic continuation to $n=1$ when computing the total von Neumann entropy, as observed in previous work \cite{IsingQuench}. It is also the only contribution to our formulae that is accessible from perturbation theory in the quench parameter \cite{DeQuench1,DeQuench2,DeQuench3,DeQuench4,DeQuench5,DeQuench6}. We can then also say that on physical grounds $\tilde{\tau}'_1=1$ and $Z_1(0)=1$ even if we can not show this analytically from our formula for $Z_n(0)$.

Because of the analytic continuation subtleties above, it makes sense to consider ratios of correlators where only the oscillatory part remains. In such cases the analytic continuation to $n\rightarrow 1$ can also be obtained. 
For example, we find that the only difference between both one-point functions (\ref{formula}) and (\ref{formula2}) at this order at least, is confined to the oscillatory terms and we have that the ratio
\beqa  
R_n:= \frac{Z_n(1)}{Z_n(0)}=\varepsilon^{\frac{1}{8n}}\frac{_n\bra{\Omega}| \mathcal{T}_\mu(0, t) |{\Omega}\ket_n}{_n\bra{\Omega}| \mathcal{T}(0, t) |{\Omega}\ket_n}=  \varepsilon^{\frac{1}{8n}}\frac{\tilde{\tau}_n'^\mu}{\tilde{\tau}_n'} \exp \left[- \frac{\alpha}{16\sqrt{\pi} n \cos^2\frac{\pi}{4 n}} \frac{\cos\!\left (2mt - \frac{\pi}{4} \right )}{\left ( m t\right )^{3/2}}\right]\,,
%\\
%&=& \varepsilon^{\frac{1}{8n}}\frac{\tilde{\tau}_n^\mu}{\tilde{\tau}_n} \exp \left[-2A \sin^2\frac{\pi}{2n}- \frac{\alpha}{16\sqrt{\pi} n \cos^2\frac{\pi}{4 n}} \frac{\cos\!\left (2mt - \frac{\pi}{4} \right )}{\left ( m t\right )^{3/2}}\right]\nonumber\,.
\label{ratio}
\eeqa 
and, with the caveats discussed above we can write that
\beqa  
R_1= \varepsilon^{\frac{1}{8}} \bra \tilde{0}|\mu(0,0)|\tilde{0} \ket \exp \left[- \frac{\alpha}{8\sqrt{\pi}} \frac{\cos\!\left (2mt - \frac{\pi}{4} \right )}{\left ( m t\right )^{3/2}}\right]\,.
\label{ratio1}
\eeqa 
Similarly,
\beqa  
F_n:= \frac{Z_n(1)}{(Z_1(1))^n} &=& \varepsilon^{-\frac{1}{12}\left(n-\frac{1}{n}\right)}\frac{_n\bra{\Omega}| \mathcal{T}_\mu(0, t) |{\Omega}\ket_n}{_n\bra{\Omega}| \mu(0, t) |{\Omega}\ket_n^n} \nonumber\\
&=&  \varepsilon^{-\frac{1}{12}\left(n-\frac{1}{n}\right)} \frac{\tilde{\tau}_n'^\mu}{\tilde{\tau}_1'^\mu}\exp{\left[-\frac{\alpha}{8 \sqrt{\pi }} \left(\frac{1}{ n \sin^2\frac{\pi }{2 n}}-n \right) \frac{\cos \left(2 m t-\frac{\pi}{4}\right)}   {(mt)^{3/2}}\right]}\,,
%\\
%&=& \varepsilon^{\frac{1}{8n}}\frac{\tilde{\tau}_n^\mu}{\tilde{\tau}_n} \exp \left[-2A \sin^2\frac{\pi}{2n}- \frac{\alpha}{16\sqrt{\pi} n \cos^2\frac{\pi}{4 n}} \frac{\cos\!\left (2mt - \frac{\pi}{4} \right )}{\left ( m t\right )^{3/2}}\right]\nonumber\,.
\label{ratio2}
\eeqa 
with $F_1=1$ as expected. 

\subsection{Symmetry Resolved Entanglement Entropy}
In the presence of $\mathbb{Z}_2$-symmetry, the 
SREE is given in terms of the discrete Fourier transform of the moments $Z_n(0), Z_n(1)$ defined in \eqref{t_SE}.
More precisely, there are two symmetry resolved partition functions, given by
\beq 
\mathcal{Z}_n(\pm)=\frac{1}{2}(Z_n(0)\pm Z_n(1))\,,
\eeq 
and the  corresponding symmetry resolved R\'enyi entropies are given by
\beqa 
S_n(\pm)=\frac{1}{1-n}\log\frac{\mathcal{Z}_n(\pm)}{(\mathcal{Z}_1(\pm))^n}&=&-\log 2+ \frac{1}{1-n} \log \left [ \frac{Z_n(0) \pm Z_n(1)}{(Z_1(0) \pm Z_1(1))^n}  \right ]\nonumber\\
&=& S_n-\log 2 + \frac{1}{1-n}\log\left[\frac{1\pm R_n}{(1\pm R_1)^n}\right]\,,
\label{ver1}
\eeqa 
where $S_n$ is the total R\'enyi entropy given by $\frac{1}{1-n}\log \frac{Z_n(0)}{Z_1(0)^n}$ and we know from \cite{IsingQuench} that this quantity grows linearly with time, that is, here we must assume that $Z_1(0)=1$ and take $Z_n(0)$ to be given by (\ref{formula2}), up to a power of $\varepsilon$. Since $\Delta_n^\mu>\Delta_n$ it is possible to argue that for $\varepsilon\ll 1$ the leading contribution to the expansion of the logarithm is
\cite{Horvath:2020vzs}  as
\begin{equation}\begin{split}
    S_n(\pm) = S_n - \log 2 \pm \frac{R_n}{1-n}  + O(\varepsilon^{\frac{1}{4n}},\varepsilon^{\frac{1}{8}})\,.
\end{split}\end{equation}
As seen in the previous subsection $R_n\, \propto \, \varepsilon^{\frac{1}{8 n}}$. 
For $n>1$ this term is the leading correction to the $\varepsilon$ expansion. However, the precise value of $n$ is important in order to determine next-to-leading order corrections. For example, if $n=2$ the term $ 2 R_1$ resulting from expanding the denominator in the logarithm, is of the same order in $\varepsilon$ as $R_2^2$. A key result of this expansion is that for large times, the SREEs grow linearly in $mt$ at the same rate as the total entropy $S_n$. Linear growth in time is expected from existing work which employs the well-known quasiparticle pair picture of entanglement \cite{quench3,quench2,AC} generalised to symmetry resolved measures \cite{Parez:2020vsp,Parez:2021pgq,Parez:2022xur}.

As long as we can argue that the term $\frac{R_n}{1-n}$ is subleading compared to the total entropy, we can also state that the SREEs display the property of equipartition. As we know \cite{GS,german3,ourreview} equipartition only holds in certain limits and it has been observed to hold even out of equilibrium. For example in \cite{Parez:2020vsp} it was noted that equipartition applies when the difference between the charge and its average is small compared to system size. In our case, $R_n$ depends on many parameters ($n$, $\alpha$, $\varepsilon$) and equipartition is found under various sets of conditions. For example, since $R_n$ scales with $\varepsilon^{\frac{1}{8n}}$
we have that $R_n$ is small for $\varepsilon\ll 1$ and intermediate values of $n$. If $n$ is large $\varepsilon^{\frac{1}{8n}}\rightarrow 1$. However, since there is also division by $1-n$ we still need the term $(1-n)^{-1} R_n$ to be very small for large $n$. 

\fed{Notice that the same quantities $S_n(\pm)$ may be expressed in terms of the functions $R_n$ and $F_n$ defined in \eqref{ratio1} by re-writing (\ref{ver1}) as
\beqa 
 \log \left [ \frac{Z_n(0) \pm Z_n(1)}{(Z_1(0) \pm Z_1(1))^n}  \right ]=\log F_n+ \log\left[\frac{R_n^{-1}\pm 1}{(R_1^{-1}\pm 1)^n}\right]= \log F_n- \log\frac{R_n}{R_1^n}+\log\left[\frac{1\pm R_n}{(1\pm R_1)^n}\right]\,,
\eeqa 
which by comparison implies that the total entropy can also be written as
\beq 
S_n=\log 2- \frac{1}{1-n} \log\frac{R_n}{F_n R_1^n}\,.
\eeq 
}
\section{The Quasiparticle Picture}

As already mentioned, the linear growth of the SREE as a function of $mt$ is expected from the quasiparticle picture of entanglement evolution.
In this section, we present the quasiparticle predictions for the moments $Z_n(0)$ and $Z_n(1)$ in Eq.~\eqref{t_SE} and compare them with the results~\eqref{formula} and \eqref{formula2} of the form factor expansion.

According to the quasiparticle picture (QPP)~\cite{EEquench,AC,AC2,PCAg}, the extensive contribution to the entanglement entropies is given by pairs of quasiparticles which are shared between the subsystem and its complement (in our case, between the $x<0$ and the $x>0$ half-space regions).
Indeed, from Eq.~\eqref{psio} we see that the initial state in the quench under study is a coherent state with a pair structure, in which only excitations with opposite momenta are entangled.
As we let the state evolve, these pairs of quasiparticles spread ballistically with opposite velocity $v(\theta) = \pm m \tanh \theta$.
In half-space, at time $t$ the only quasiparticles which contribute to entanglement are those in the light-cone $x < 2 t$, whose corresponding pair is in the opposite half.
The quasiparticle picture ansatz then posits that each of the shared quasiparticles contributes to the entropies with their Yang-Yang entropy~\cite{Alba:2017kdq,BFPC_2018}.
Importantly, this ansatz is non-perturbative; despite this, as we will show, it only captures the linear growth while it fails to predict the oscillations.

Originally introduced to compute the entanglement entropy, the QPP has been extended to other quantities, such as certain correlation functions~\cite{CEF}, logarithmic negativity~\cite{CTC,Alba:2018hie}, full counting statistics~\cite{Groha:2018bah,Horvath:2023dgd,Bertini:2022srv} and operator entanglement~\cite{Dubail:2016xht,Rath:2022qif}. 
In particular, in Ref.~\cite{Parez:2020vsp,Parez:2021pgq} the QQP was used to obtain the $U(1)$ symmetry resolved entanglement entropy.
In the present paper, to study the $\Z_2$ symmetry resolution we restrict the $U(1)$ result to the $\Z_2$ subgroup.
In the half-space $x>0$, the QPP prediction for the charged moment is then 
\begin{equation}\label{eq:QPans}\begin{split}
    \log Z_n(a) &= \int_{0}^{+\infty} \frac{\dd k}{2\pi}\, 2 v(k) t\, \log\left [ (-1)^a  \xi(k)^n + \left ( 1- \xi(k) \right )^n \right ]\\
    &= 2 m t \int_{0}^{+\infty} \frac{\dd \theta}{2\pi} \sinh \theta\, \log\left [ (-1)^a  \xi(\theta)^n + \left ( 1- \xi(\theta) \right )^n \right ], \quad \text{with } a = 0, 1,
\end{split}\end{equation}
where $v(\theta) = m \tanh \theta$ is the velocity of the excitations with rapidity $\theta$ and $\xi(\theta)$ is their occupation number.
In the expression~\eqref{eq:QPans}, inside the integral we can recognise an ``occupation'' contribution proportional to $\xi(\theta)$ and a ``vacancy'' contribution proportional to $(1-\xi(\theta))$.
Since the fermionic modes are odd under the $\Z_2$ symmetry, we see that in the charged moment $Z_n(1)$ the occupation contribution is the one carrying the charge $(-1)$, while the vacancy one is identical between $Z_n(0)$ and $Z_n(1)$.

For squeezed initial states of the form~\eqref{psio}, the occupation number is related to the function $\widehat{K}(\theta)$ in Eq.~\eqref{KK} as
\begin{equation}\label{eq:occupation}
    \xi(\theta) = \frac{\widehat{K}(\theta)^2}{1+\widehat{K}(\theta)^2} \approx \widehat{K}(\theta)^2 + \ldots ,
\end{equation}
which shows that our perturbative expansion in $\widehat{K}(\theta)$ corresponds to an expansion in small density of excitations.
Interestingly, since $\widehat{K}(\theta) \approx \xi(\theta)^{1/2}$, the first order oscillatory term in Eqs.~\eqref{formula} and \eqref{formula2} is non-analytical in the density $\xi(\theta)$; as a consequence, since the QPP ansatz~\eqref{eq:QPans} is analytic in $\xi(\theta)$ it can never reproduce this leading oscillatory term.

In order to compare Eq.~\eqref{eq:QPans} with the form factor results \eqref{formula} and \eqref{formula2} of the present paper and of \cite{IsingQuench}, we expand them in small $\widehat{K}(\theta)$ using equation  \eqref{eq:occupation}.
Notice that, while for $Z_n(0)$ the comparison with the QPP result was already performed in \cite{IsingQuench}, in that paper both the quasiparticle and the form factor results were further expanded in the parameter $\alpha$ of \eqref{alpha}; in this paper we instead directly compare the leading orders in $\widehat{K}(\theta)$.
Substituting \eqref{eq:occupation} in \eqref{eq:QPans} and expanding, we find for the charged moments
\begin{equation}\label{eq:QPans2}\begin{split}
    \log Z_n(a) &\approx 2mt\int_0^{+\infty} \frac{\dd\theta}{2\pi}\, \sinh(\theta)\, \log\!\left [ (-1)^a\left ( \widehat{K}(\theta)^2 + \ldots \right )^n + \left ( 1 - \widehat{K}(\theta)^2 + \ldots \right )^n \right ]\\
    &\approx 2mt\int_0^{+\infty}\, \frac{\dd\theta}{2\pi}\, \sinh(\theta)\, \log \!\left [ 1 - n \widehat{K}(\theta)^2 + \ldots \right ] \\
    &\approx - 2 n m t \int_{0}^{+\infty} \frac{\dd \theta}{2\pi}\, \widehat{K}(\theta)^2\, \sinh(\theta) + O(K^2) = - \frac{n\Gamma}{2}\, m t  + O(K^2)\,, 
\end{split}\end{equation}
where $\Gamma$ is precisely the rate of growth obtained in \eqref{G}, correctly reproducing the linear growth term in equations \eqref{formula} and \eqref{formula2}, as expected.
Before concluding this section, we point out that the small density approximation~\eqref{eq:QPans2} gives an additional perspective on the fact that the growth rate $\Gamma$ is at leading order identical for both moments $Z_n(0)$ and $Z_n(1)$.
Indeed, in \eqref{eq:QPans2} we see that the leading order in $\widehat{K}(\theta)$ is entirely given by the vacancy contribution $1- \xi(\theta) \approx 1 - \widehat{K}(\theta)^2$ of \eqref{eq:QPans}.
As we mentioned previously, the dependence on the charge $(-1)^a$ is entirely given by the occupation contribution, while the vacancy contribution is the same for both moments, yielding the same result at leading order.

\section{Conclusion}
\label{5}
In this paper we have generalised the computation of the one-point function of the branch point twist field $\TT$ presented in \cite{IsingQuench} and of the magnetisation or order parameter carried out in \cite{SE} to the composite twist field $\TT_\mu$. The latter plays a prominent role in the computation of the symmetry resolved entanglement entropy of the Ising field theory, where resolution is with respect to the $\mathbb{Z}_2$ symmetry of the model.  We study the time evolution of the one-point function of $\TT_\mu$ after a mass quench. This is a global quench, in the sense that it is associated with the sudden change of a global parameter and, importantly, it is a quench that preserves the $\mathbb{Z}_2$ symmetry of the theory. This mass quench is fully characterised by a post-quench (boundary-like) state in which pairs of quasiparticles of opposite momenta are created in the entanglement region and become entangled with quasiparticles in the complementary region.

We find that both the entanglement entropy and the symmetry resolved entanglement entropy grow linearly in time, with the same growth rate. Up to a factor $n$ which accounts for the replicas, this is also the same decay rate of the magnetisation, thus a characteristic of all symmetry fields in the theory. \fed{In particular, the one-point functions of all fields considered in this paper, $\TT$, $\TT_\mu$ and $\mu$, tend to zero as $t \rightarrow \infty$. This is a distinct feature of the quench dynamics of symmetry fields, as recently shown in \cite{Salvo_2025}}. Furthermore, the SREE exhibits equipartition with respect to the symmetry sector, up to corrections that are subleading for small UV cut-off and/or large $n$.  Similarly to \cite{IsingQuench} we could not find a well-defined $n\to 1$ limit of our entropies. A possible explanation is that this limit is not compatible with the regularisation adopted. In \cite{Salvo_2025} a finite-size regularisation was used. It is possible that in this way the von Neumann entropies can be effectively obtained.

It would be very interesting to extend this kind of computation to other theories, especially interacting ones.
\medskip

\noindent {\bf Acknowledgements:} 
Federico Rottoli is grateful to V. Alba and C. Rylands for useful discussions on the quasiparticle picture.
  Michele Mazzoni acknowledges PhD funding under the EPSRC Mathematical Sciences Doctoral Training Partnership EP/W524104/1. Fabio Sailis is funded through a PhD studentship from the School of Science and Technology of City St George's, University of London.  Federico Rottoli thanks the Department of Mathematics at City, University of London for hospitality during a one-month visit in 2024 when this project was initiated.
  Federico Rottoli acknowledges support from the project ``Artificially devised many-body quantum dynamics in low dimensions - ManyQLowD'' funded by the MIUR Progetti di Ricerca di Rilevante Interesse Nazionale (PRIN) Bando 2022 - Grant 2022R35ZBF.

\bibliography{Ref}
\end{document}